\documentclass[final,leqno]{siamltex}

\usepackage[utf8]{inputenc}
\usepackage[english]{babel}
\usepackage{amsmath, amssymb}

\usepackage[hang]{caption}
\usepackage{graphicx}
\usepackage{subfig}
\usepackage{nicefrac}
\usepackage{xcolor}
\usepackage{tikz}

\usepackage{hyperref}
\usepackage{bookmark}
\hypersetup{
  pdftitle    = {Designing Illumination Lenses and Mirrors by the Numerical Solution of Monge-Amp\`ere Equations},
  pdfauthor   = {Kolja Brix, Yasemin Hafizogullari, and Andreas Platen},
  pdfkeywords = {Inverse reflector problem, inverse refractor problem, elliptic Monge-Amp\`ere equation, B-spline collocation method, Picard-type iteration},
  pdfborder={0 0 0.5}
}

\newcommand{\R}{\ensuremath{\mathbb{R}}}
\newcommand{\Z}{\ensuremath{\mathbb{Z}}}
\newcommand{\N}{\ensuremath{\mathbb{N}}}
\newcommand{\DD}{\ensuremath{\textnormal{D}}}
\newcommand{\dd}{\ensuremath{\textnormal{d}}}
\newcommand{\GG}{\ensuremath{\nabla}}
\newcommand{\GGG}{\ensuremath{\widehat{\nabla}}}

\newcommand{\cA}{\ensuremath{\mathcal{A}}}
\newcommand{\cI}{\ensuremath{\mathcal{I}}}
\newcommand{\cM}{\ensuremath{\mathcal{M}}}
\newcommand{\cN}{\ensuremath{\mathcal{N}}}
\newcommand{\cO}{\ensuremath{\mathcal{O}}}
\newcommand{\cS}{\ensuremath{\mathcal{S}}}

\newcommand{\abs}[1]{\ensuremath{|#1|}}
\newcommand{\abslr}[1]{\ensuremath{\left|#1\right|}}
\renewcommand{\vec}[1]{\ensuremath{\mathbf{#1}}}

\author{Kolja Brix\footnotemark[4] \and
Yasemin Hafizogullari\footnotemark[4] \and
Andreas Platen\footnotemark[4]}

\date{\today}
\title{Designing Illumination Lenses and Mirrors by the Numerical Solution of Monge--Amp\`ere Equations}

\begin{document}

\maketitle

\renewcommand{\footnotesep}{1.0em}
\renewcommand{\thefootnote}{\fnsymbol{footnote}}

\footnotetext[4]{Institut f{\"u}r Geometrie und Praktische Mathematik, RWTH Aachen, Templergraben 55, 52056 Aachen, Germany
(\texttt{brix@igpm.rwth-aachen.de},
\texttt{yasemin.hafizogullari@rwth-aachen.de},
\texttt{andreas.platen@rwth-aachen.de}),
\url{http://www.igpm.rwth-aachen.de}.}

\footnotetext{Copyright \textcopyright\ 2015 Optical Society of America. One print or electronic copy may be made for personal use only. Systematic reproduction and distribution, duplication of any material in this paper for a fee or for commercial purposes, or modifications of the content of this paper are prohibited.}

\renewcommand{\thefootnote}{\arabic{footnote}}

\begin{abstract}
We consider the inverse refractor and the inverse reflector problem. The task is to design a free-form lens or a free-form mirror that, when illuminated by a point light source, produces a given illumination pattern on a target.
Both problems can be modeled by strongly nonlinear second-order partial differential equations of Monge--Amp\`ere type.
In [Math. Models Methods Appl. Sci. 25 (2015), pp. 803--837, DOI: \href{http://dx.doi.org/10.1142/S0218202515500190}{10.1142/S0218202515500190}] the authors have proposed a B-spline collocation method which has been applied to the inverse reflector problem. Now this approach is extended to the inverse refractor problem.
We explain in depth the collocation method and how to handle boundary conditions and constraints.
The paper concludes with numerical results of refracting and reflecting optical surfaces and their verification via ray tracing.
\end{abstract}



%
%
{\footnotesize
\noindent\hspace{5mm}\textbf{OCIS.}\,
  (000.4430) Numerical approximation and analysis,
 %
  (080.1753) Computation methods,
  (080.4225) Nonspherical lens design,
  (080.4228) Nonspherical mirror surfaces,
  (080.4298) Nonimaging optics,
 %
  (100.3190) Inverse problems
}
%


\pagestyle{myheadings}
\thispagestyle{plain}

\markboth{BRIX, HAFIZOGULLARI, AND PLATEN}{DESIGNING ILLUMINATION LENSES AND MIRRORS}


\section{Introduction}

\begin{figure}
  \centering
  \includegraphics[width=0.6\linewidth]{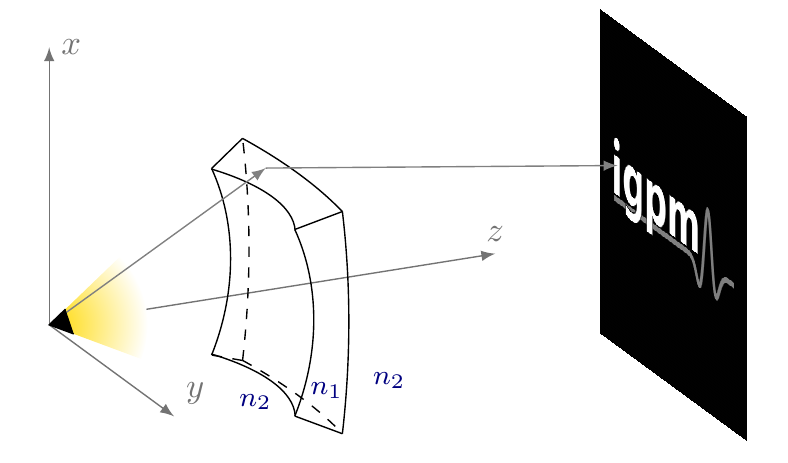}
  \caption{Setting of the refractor problem. The index of refraction of the lens material is $n_1$, while the surrounding has the refractive index $n_2$.}
  \label{fig:parametrization}
\end{figure}

Both problems, the inverse refractor and the inverse reflector problem, from illumination optics can be formulated in the following framework:
Let a point-shaped light source and a target area be given, e.g. a wall. Then we would like to construct an apparatus that projects a prescribed illumination pattern, e.g. an image or a logo, onto the target.
Since we aim for maximizing the efficiency, we would like to construct our optical device in such a way that, neglecting losses, it redirects all light emitted by the light source to the target.
We focus our attention to the design of such an optical system in the simple case that it either consists of a single free-form lens or of a single free-form mirror, see Figure~\ref{fig:parametrization} for an illustration of the former case.
Our goal is now to compute the shape of the optically active surfaces, modeled as \emph{free-form surfaces}, such that the desired light intensity distribution is generated on the target.
Since these problems from illumination optics from the mathematical point of view conceptually fall into the class of inverse problems, they are also called \emph{inverse reflector problem} and \emph{inverse refractor problem}, respectively.
In particular, since the size of the optical system is comparable to that of the projected image, we address the case of the near field problems.

There is a variety of technical applications of such optical systems, e.g. spotlights with prescribed illumination patterns used in street lamps or car headlamps, see e.g.~\cite{BBW+2012,BBL+2011,SS2012}.

The authors present in~\cite{BHP2015} a solution method for the inverse reflector problem via numerically solving a strongly nonlinear second-order \emph{partial differential equation (PDE) of Monge--Amp\`ere type}.
Due to the high potential of this approach we now extend this method to the case of illumination lenses.

This paper is organized as follows:
Since the reflector problem has been discussed in detail in~\cite{BHP2015} we mainly focus on the refractor problem. We start with the state of the art for its solution in Section~\ref{sec:state_of_the_art}. Then, we formulate the problem via a partial differential equation of Monge--Amp\`ere type which we discuss in Section~\ref{sec:refractor_problem} for the construction of a refractor. For completeness we also give the Monge--Amp\`ere formulation for the reflector problem in Section~\ref{sec:reflector_problem}.
Next, the numerical method is explained in Section~\ref{sec:collocation}.
Since this type of optical design problem raises many difficulties in the solution process we discuss in Section~\ref{sec:MA_in_optics} how these can be resolved.
Finally, in Section~\ref{sec:results} we look at numerical results for the inverse reflector and refractor problems and end this paper in Section~\ref{sec:outlook} with our conclusions.


\section{State of the art}\label{sec:state_of_the_art}

In this section we discuss the methods available for the solution of the inverse design problems in nonimaging optics, see the monographies by Chaves~\cite{Chaves2008} and by Winston, Mi\~nano and Ben\'itez~\cite{WMB2005} for an introduction to nonimaging optics and the paper by Patow and Pueyo~\cite{PP2005} for a survey article on inverse surface design from the graphics community's point of view.
For a detailed survey of solution techniques for the inverse reflector problem, we refer the reader to \cite[Section 2]{BHP2015}.

Focusing on the inverse refractor problem, in the paper by Wester an B\"auerle~\cite{WB2013} there is a list of approaches, a discussion on practical problems, e.g. extended sources and Fresnel losses, and examples with LED lighting, e.g. a lens for automotive fog light and a lens producing a logo.

In the rest of this section, we first give a short overview of other solution techniques in Section~\ref{sect:SotA_NoPDEapproach} and then focus on methods based on PDEs in Section~\ref{sect:SotA_PDEapproach}, which is also our problem formulation of choice.
Finally, we discuss some advanced topics in Section~\ref{sect:SotA_AdvancedTopics} and draw our conclusions in Section~\ref{sect:SotA_Conclusion}.

\subsection{Approaches for the solution of inverse problems in nonimaging optics not based on a PDE}\label{sect:SotA_NoPDEapproach}

We distinguish three different groups of techniques for the solution of inverse problems in nonimaging optics, which are not based on a PDE: there are methods resorting from optimization techniques, others built from Cartesian ovals and a third group of methods which are geometrical constructions.

\paragraph{Optimization approaches}

There are methods for the design of optical surfaces, which are based on optimization techniques, see e.g. \cite{RW2007,WWL+2013}.
Starting from an initial guess, the outline of the iterative optimization process for the determination of the optical surfaces is as follows:
First, the current approximation of the optical surfaces is validated by ray tracing.
In a second step, using an objective function, which is often closely related to the Euclidean norm, the resulting irradiance distribution is compared to the desired one and a correction of the optical surfaces is determined.
The process ends, when a suitable quality criterion is fulfilled, otherwise these two steps are repeated.

The advantage of this method is that it is very flexible. However, optimization procedures are very costly because of the repeated application of the ray tracing and it is unclear if the iterative methods converge at all.

\paragraph{Cartesian ovals methods}

Cartesian ovals are curves of fourth order in the plane. They can be associated with two foci such that light emitted at one focus is collected at the other focus. Here the Cartesian oval coincides with the interface of two optical media with different refractive indices.
Cartesian ovals can be extended to surfaces in 3d with the same property.
By combining clippings of several of these surfaces in an iterative procedure a new segmented surface can be constructed that approximates the solution.
This strategy has first been developed by Kochengin and Oliker~\cite{KO1997,KO1998} for the construction of solutions for the inverse reflector problem. Later this has been extended to the inverse refractor problem~\cite{MSB2011} using Cartesian ovals, see also \cite[Section 2]{Gutierrez2014} for some theoretical background, and for a collimated light beam instead of a point light source~\cite{Oliker2011} using hyperboloids.

Although this technique has the advantage to permit the construction of continuous but non-differentiable surfaces~\cite{MSB2011}, the number of clippings $K$ required grows linearly in the number of pixels in the image. For example, using ellipsoids of revolution for the construction of a mirror with accuracy $\gamma>0$, the complexity of the method scales like $\cO(\frac{K^4}{\gamma} \log{\frac{K^2}{\gamma}})$, see \cite{KO2003}, such that it quickly becomes infeasible for higher resolutions.

\paragraph{Geometric construction methods}

Reflective and refractive free-form surfaces can also be designed by geometric approaches.
Probably the most famous of these techniques is the simultaneous multiple surfaces (SMS) method extending the ideas of Cartesian-oval methods, see e.g. \cite[Chapter 8]{WMB2005} and \cite{BMB+2004, MBL+2009} and the references therein.
The main idea of the SMS method is the simultaneous construction of two optical surfaces, e.g. both surfaces of a lens, which permits to couple two prescribed incoming wave fronts, e.g. coming from two point light sources, with two prescribed outgoing wave fronts.
While in its 2D version, the method is used to design rotationally symmetric optical surfaces, in a 3D variant it is also capable to construct free-form optical surfaces.
However, the authors could not find any hint on the computational costs in the literature but conjecture that this scheme is expensive especially for complex target illumination patterns.

\subsection{Solution techniques via PDE approaches}\label{sect:SotA_PDEapproach}

In several publications for the inverse refractor problem a PDE is derived, whose solution models the desired optical free-form surfaces, see e.g. \cite{SS2012,Gutierrez2014,GH2014,Oliker2014,WXL+2013,RM2001,RM2002,DLZ+2008}. In these approaches usually the low wavelength limit is assumed to hold, i.e. the problems are formulated using the geometrical optics approximation.

Some examples for the inverse refractor problem with a more complex target illumination pattern are shown in \cite{SS2012,WXL+2013,RM2001,RM2002}. However, in all four articles the descriptions and discussions of the numerical methods are incomplete. To the best of the authors' knowledge the solution method is not fully documented in the literature.

While we consider the case of a point light source, an interesting and closely related problem is shaping the irradiance distribution of a collimated light beam, see e.g. \cite{Oliker2011,Oliker2014} for the theory including some results on existence and uniqueness of solutions.

We refer the reader to the monography by Guti\'errez~\cite{Gutierrez2001} for a general overview of Monge--Amp\`ere-type equations.

Since we are looking for an optical surface which redirects light coming from a source onto a target, one can model this problem in terms of optimal transportation.

\paragraph{Optimal transport}

There are also methods which are based on a problem of optimal transport which leads to Monge--Amp\`ere-type equations, see e.g. \cite{BBW+2012,BBL+2011,ORW2013}. First the ray mapping, i.e. the mapping of the incoming light rays onto the points at the target, is computed via an optimal transport approach. At this point the optical surface is still unknown but in a next step it is constructed from the knowledge of the target coordinates for each incoming light ray. In 1998 Parkyn~\cite{Parkyn1998} already described a very similar procedure.

\subsection{Advanced topics}
\label{sect:SotA_AdvancedTopics}

In the current formulation of the problem only one single idealized point light source has been used. An extension to multiple point light sources is discussed by Lin~\cite{Lin2012} where the optical refractors are determined from those calculated for single point light sources by a weighted least-squares approach. More techniques for the case of extended light sources can be found in the papers by Bortz and Shatz~\cite{BS2006} and Wester et al.~\cite{WMV+2014}.

In particular for the refractor problem, some energy is lost for the illumination of the target because of internal reflections in the lens material.
A theoretical discussion of these Fresnel losses can be found in the publications by Guti\'errez~\cite[Section 5.13]{Gutierrez2014} and Guti\'errez and Mawi~\cite{GM2013}.
In \cite{BBW+2012,BBL+2011} the losses are minimized by free-form shaping of both refractive surfaces of the lens.

\subsection{Conclusion}
\label{sect:SotA_Conclusion}

Our approach is motivated by the fact that even for the special case of a single point light source and the computation of just one surface of the lens we could not find any fully detailed method in the literature which can produce complex illumination patterns on the target area.
From the authors' point of view, the most promising approach is the one by solving a
PDE of Monge--Amp\`ere type.


\section{The inverse refractor problem}\label{sec:refractor_problem}

This section is devoted to the formulation of the Monge--Amp\`ere equation that models the near field refractor problem as given in the paper by Guti\'errez and Huang~\cite{GH2014}.
Since the full theory is a bit involved, we restrict ourselves to a summary of the most important aspects and refer the reader to \cite[Appendix A]{GH2014} and the paper by Karakhanyan and Wang~\cite{KW2010} for the details.
Our notation also follows these sources.

We now proceed as follows: At first, we fix the geometric setting and the implicit definition of the refracting and the target surfaces in Section~\ref{subsec:geo_setting}. Then we apply Snell's law of refraction in Section~\ref{subsec:snell} and follow the path of the light ray in Section~\ref{subsec:light_path}. Finally, in Section~\ref{subsec:MA} we obtain the desired equation of Monge--Amp\`ere type.

\subsection{The Geometric Setting}\label{subsec:geo_setting}

Since a lens has two surfaces we need to design both of them.
For simplicity we choose a spheric inner surface, i.e. the surface which faces the light source is a subset of a sphere with center at the position of the light source. Thus there is no refraction of the incoming light at this interface, the inner surface is optically inactive.

It remains to compute the shape of the outer surface facing the target area.
To that end let us define the quotient $\kappa=\frac{n_2}{n_1}$ of the refractive indices of the lens material $n_1$ and the environment $n_2$.
We assume that the light source illuminates a non-empty subset $U$ of the northern hemisphere of the unit sphere $\cS^2 \subset \R^3$. The third component of an incoming light ray with direction $\vec{x} = (x_1,x_2,x_3)^T\in U$ is then given as $x_3=\sqrt{1-x_1^2-x_2^2}$. Thus we define $\vec{x}':=(x_1,x_2)^T$ and parametrize our outer lens surface by the distance function $\rho:=\rho(\vec{x}')$, i.e. the surface is given as $\Gamma:=\{\rho(\vec{x}')\vec{x}\,:\,\vec{x}\in U\}$.

The target $\Sigma$ is defined as a subset of a hypersurface implicitly given by the zero level set of a continuously differentiable function $\psi$ via
\begin{equation}\label{eq:sigma}
\Sigma\subset\{\vec{z}\in\R^3\,:\,\psi(\vec{z})=0\}.
\end{equation}
Note that for the numerical solution procedure in the Newton-type method we require that $\psi$ is twice continuously differentiable.
While in general much more complicated situations are supported \cite{GH2014}, for simplicity we restrict ourselves to the case where the target $\Sigma$ is on a shifted $x$-$y$-plane such that $\psi(\vec{z}):=z_3-\gamma$ for a shift $\gamma>0$.

To model the \emph{luminous intensity} of the source we define the density function $f:U\rightarrow \R^+$, where $\R^+:=\{x\in\R\,:\,x>0\}$. The corresponding density function for the desired illumination pattern on the target $\Sigma$ is denoted by $g:\Sigma\rightarrow \R^+$.
Since we want to redirect all incoming light onto the target the density functions need to fulfill the \emph{energy conservation} condition
\begin{align}\label{eq:energy_conservation}
  \int_{U} f\,\dd S= \int_{\Sigma} g \, \dd S.
\end{align}
Note that for simplicity we neglect the loss of reflected light intensity. For a more complicated derivation of a Monge--Amp\`ere-type equation for the refractor problem taking losses into account see \cite{GM2013}.

\subsection{Snell's law of refraction}\label{subsec:snell}

According to Snell's law of refraction in vectorial notation (see e.g. \cite[Chapter 4.4]{Hecht2013} or \cite[Chapter 12]{Chaves2008}), the direction of the light ray after refraction at the point $\rho\vec{x}$ is $\vec{y} = \frac{1}{\kappa} ( \vec{x} - \Phi(\vec{x} \cdot \nu) \nu) \in \cS^2$,
where
$\Phi(s) := s - \sqrt{\kappa^2+s^2-1}$
and $\nu$ is the outer unit normal on $\Gamma$ defined as a function on $U$.

As detailed in \cite[(2.15)]{KW2010}, for the outer normal unit vector at $\vec{x} \in U$ we find
\begin{equation} \label{eq:ONUV}
\nu = \frac{- \GGG \rho(\vec{x}') + \vec{x} (\rho(\vec{x}') + \GG \rho(\vec{x}') \cdot \vec{x}')}{\sqrt{\rho^2(\vec{x}') + \abs{\GG \rho(\vec{x}')}^2 - (\GG \rho(\vec{x}') \cdot \vec{x}')^2}},
\end{equation}
where $\GG f$ denotes the gradient of a function $f$ and $\GGG\rho(\vec{x}'):=(\GG\rho(\vec{x}'),0)\in\R^3$.
To ease notation, we define the utility function $G$
which represents the denominator in \eqref{eq:ONUV}, i.e.
\begin{equation}\label{eq:defG}
G(\vec{x}', u, \vec{p}) := \sqrt{u^2 + \abs{\vec{p}}^2 - (\vec{p} \cdot \vec{x}')^2 }.
\end{equation}

\subsection{Following the light ray}\label{subsec:light_path}

Next, we consider the line which contains the light ray after refraction, defined by the point $\rho \vec{x}$ and the direction vector $\vec{y}$.
We now turn to finding the point $\vec{z} = (z_1,z_2,z_3)^T$ where the refracted light ray hits the target $\Sigma$.
In order to determine the third component $z_3$ of $\vec{z}$, we first define the utility point $\vec{w}=(w_1, w_2, 0)^T$ as the intersection point of this line with the plane $\{ \vec{x} \in \R^3: x_3 = 0 \}$ which is given as
\begin{equation*}
\vec{w} = \rho(\vec{x}') \vec{x} + d_0 \vec{y}
\end{equation*}
for a $d_0\in\R$. For a proof of the existence of $\vec{w}$ see \cite[Appendix A.2]{GH2014}.
Using \eqref{eq:ONUV}, we confirm that
\begin{equation}\label{eq:defw}
\vec{w} = F(\vec{x}', \rho(\vec{x}'), \GG \rho(\vec{x}')) \GGG (\rho^2),
\end{equation}
where the utility function $F$ is given by
\begin{equation}\label{eq:defF}
  F(\vec{x}', u, \vec{p}) := \frac{1}{2} \frac{\Phi(\nicefrac{u}{G(\vec{x}', u, \vec{p})})}{-G(\vec{x}', u, \vec{p})+(u+\vec{p} \cdot \vec{x}') \Phi(\nicefrac{u}{G(\vec{x}', u, \vec{p})})}.
\end{equation}
After refraction at the point $\rho\vec{x}$ the light ray hits the target $\Sigma$ at point $\vec{z} = (z_1,z_2,z_3)^T$ given as
\begin{align*}
\vec{z}(\vec{x}') = \rho(\vec{x}') \vec{x} + d_1 \vec{y}
  =\rho(\vec{x}') \vec{x} + t (\vec{w} - \rho(\vec{x}') \vec{x}).
\end{align*}
From the third component we know that $t = \frac{\rho x_3-z_3}{\rho x_3}$.

We introduce the short notation $\vec{x} \otimes \vec{y} := \vec{x} \vec{y}^T$.
Let us define $F := F(\vec{x}', \rho, \GG \rho)$ and denote its partial derivatives by $\DD_{\vec{x}'}F$, $\DD_{\rho}F$ and $\DD_{\vec{p}}F$, respectively.

A lengthy computation using standard calculus and some tensor identities of Sherman--Morrison type yields
\begin{equation*}
\DD \vec{w}' = 2 \rho F \cM \DD^2 \rho + B
\end{equation*}
where
$\cM := I + \frac{1}{F} \GG \rho \otimes \DD_{\vec{p}} F$
and
$B := 2 F \GG \rho \otimes \GG \rho + \GG(\rho^2) \otimes \DD_{\vec{x}'} F + \DD_u F \GG(\rho^2) \otimes \GG \rho$.
Note that
\begin{equation*}
  \cM^{-1} = I - \frac{\GG \rho \otimes \DD_{\vec{p}} F}{F + \GG \rho \cdot \DD_{\vec{p}} F}.
\end{equation*}

In a bit more involved computation along the same lines we compute
\begin{equation*}
\DD \vec{z}'= 2 t \rho F \cM (1 - \beta (\vec{w}' - \rho \vec{x}') \cdot \widetilde{\GG} \psi) (\DD^2 \rho + \cA)
\end{equation*}
with
$\beta := ( \GG \psi \cdot (\vec{w}-\rho \vec{x}) )^{-1}$,
$\widetilde{\GG} \psi := (\psi_{p_1},\psi_{p_2})$
and
$\cA = \cA(\vec{x}', \rho, \GG \rho)$, where
\begin{align*}
\cA &:= \frac{1}{2 t \rho F} \cM^{-1} (t B + (1-t) C) \quad \text{and}\\
C &:= \DD(\rho \vec{x}') + \frac{1}{\rho x_3} (\vec{w}'- \rho \vec{x}') \otimes \GG (\rho x_3).
\end{align*}

\subsection{Monge--Amp\`ere equation}\label{subsec:MA}

The energy conservation~\eqref{eq:energy_conservation} clearly also holds if we replace $U$ with any arbitrary subset $\tilde{U}\subset U$ and $\Sigma$ with $\tilde{\Sigma}:=T(\tilde{U})\subset\Sigma$, where $T:U\to\Sigma$, $\vec{x}\mapsto\vec{z}(\vec{x}')$. By coordinate transformation this yields the identity
$\det(\DD \vec{z}) = f / (g \sqrt{1-\abs{\vec{x}'}^2})$.
Finally, we can derive the Monge--Amp\`ere equation for the refractor problem
\begin{align}\label{eq:MA_refractor}
  \det(\DD \rho + \cA) =
  \frac{f(\vec{x})}{g(\vec{z}(\vec{x}')) H},\quad\text{for}\quad\vec{x'}\in\Omega
\end{align}
where $\Omega:=\{(x_1,x_2)^T\in\R^2\,:\, (x_1,x_2,x_3)^T\in U\}$ and $H=H(\vec{x}', \rho, \GG \rho)$ is computed by
\begin{align*}
  H &:= (1-\abs{\vec{x}'}^2) \abs{\GG \psi} (2t)^2 \rho^3 (-\beta) F (F+\GG \rho \cdot \DD_{\vec{p}} F),
\end{align*}
see \cite[Appendix A]{GH2014}.

\paragraph{Existence and uniqueness of solutions}

In general, for boundary value problems with Monge--Amp\`ere equations proving well-posedness, i.e. existence and uniqueness of the solution and continuous dependency on the parameters, is a hard problem, e.g. see \cite[Section 1.4]{FGN2013} for an example of a discretized Monge--Amp\`ere equation obtained by finite differences on a grid of $4 \times 4$ cells which has $16$ different solutions.

Some theoretical results for the existence of a solution for the refractor problem under some appropriate conditions can be found in \cite{GH2014} in Theorem 5.8 for $\kappa<1$ and Theorem 6.9 for $\kappa>1$.
Additionally there are results on the uniqueness of the solution if just finitely many single points on the target are illuminated, see \cite[Theorem 5.7]{GH2014} for $\kappa<1$ and \cite[Theorem 6.8]{GH2014} for $\kappa>1$.

For proving existence and uniqueness of a solution one typically requires the equation of Monge--Amp\`ere type to be elliptic. A necessary condition is that the right-hand side of \eqref{eq:MA_refractor} is positive. For this reason we demand that $\beta <0$ or, equivalently, $\GG \psi \cdot (\vec{w}-\rho \vec{x})<0$. If this term is positive we can simply replace $\psi$ by $-\psi$.


\section{The inverse reflector problem}\label{sec:reflector_problem}

The inverse reflector problem can be modeled as a Monge--Amp\`ere-type equation very similarly to the case of the inverse refractor problem in Section~\ref{sec:refractor_problem}, see~\cite{KW2010}.

Using the same definitions and notation as in Section~\ref{sec:reflector_problem} and introducing the substitution $u:=\frac{1}{\rho}$, we first define
\begin{align*}
  t &:= 1-u\frac{z_3}{x_3}, & \tilde{a} &:= \abs{\GG u}^2 - (u- \GG u \cdot \vec{x} )^2\\
  \cN &:= I+\frac{\vec{x}\otimes\vec{x}}{x_3^2}, & \tilde{b} &:= \abs{\GG u}^2 +u^2 -(\GG u \cdot \vec{x})^2,\\
  \vec{w} &:= \frac{2}{\tilde{a}} \GGG u, \quad \text{and} &
  \vec{z} &:= \frac{1}{u}\vec{x} + t\left(\vec{w}-\frac{1}{u}\vec{x}\right).
\end{align*}
We assume that $t>0$, i.e., $\frac{x_3}{u}>z_3$, and $\GG\psi \cdot (\vec{w}-\frac{1}{u}\vec{x})>0$.
Then the Monge--Amp\`ere equation for the inverse reflector problem reads
\begin{align*}
  \det\left(\DD^2 u+\frac{\tilde{a}z_3}{2tx_3}\cN\right)
    = -\frac{(u\vec{w}-\vec{x})\cdot\GG \psi}{t^2\abs{\GG\psi}x_3^2}\cdot
      \frac{\tilde{a}^3}{4\tilde{b}}\cdot\frac{f(\vec{x})}{g(\vec{z})},
\end{align*}
see \cite{KW2010} and \cite{BHP2015} for the details.


\section{Numerical solution of partial differential equations of Monge--Am\-p\`ere type}\label{sec:collocation}

The numerical solution of strongly nonlinear second-order PDEs, including those of Monge--Amp\`ere type, is a highly active topic in current mathematical research. There are many different approaches available on the market, see the review paper by Feng, Glowinski and Neilan~\cite{FGN2013} and also \cite{BHP2015} for an overview.

However, most methods are not well-suited for all equations of Monge--Amp\`ere type such that it remains unclear if a particular method can be successfully applied to our problems.
In \cite{BHP2015} the authors propose to use a spline collocation method which turns out to provide an efficient solution strategy for Monge--Amp\`ere equations arising in the inverse reflector problem.

In Section~\ref{subsec:collocation} we explain the idea of a collocation method, which reduces the problem to finding an approximation of the solution within a finite dimensional space. Then we discuss the choice of appropriate basis functions in Section~\ref{subsec:splines}.

\subsection{Collocation method}\label{subsec:collocation}

As discretization tool for the Monge--Amp\`ere equations arising in the reflector and refractor problem, we propose a collocation method, see e.g. Bellomo et al.~\cite{BLR+2008} for examples of collocation methods applied to nonlinear problems.
Let the PDE $F(\vec{x}, u, \GG u, \DD^2 u)=0$ in $\Omega$ and constraints $G(\vec{x},u,\GG u)=0$ on $\partial\Omega$ be given.
In this setting we approximate $u$ in a finite-dimensional trial subspace of $C^2(\Omega)$, i.e.
for some finite set $\cI$ and basis functions $(B_i)_{i \in \cI} \subset C^2(\Omega)$ we choose the ansatz $\hat{u} = \sum_{i \in \cI} c_i B_i$.
Next, we only require that the PDE holds true on a collocation set $\hat{\Omega}\subset \Omega$ which contains only finitely many points.
So our approximation $\hat{u}$ of the solution of our PDE satisfies
\begin{equation}
\begin{aligned}
F(\tau, \hat{u}(\tau), \GG \hat{u}, \DD^2 \hat{u}) &= 0, && \text{for} \quad \tau \in \hat{\Omega},\\
G(\tau, \hat{u}(\tau), \GG \hat{u})                &= 0, && \text{for} \quad \tau \in \partial \hat{\Omega}.
\end{aligned}
\end{equation}
This discrete nonlinear system of equations is solved by a quasi-Newton method, which uses trust-region techniques for ensuring global convergence of the method, see Chapter~4.2.1 in \cite{BHP2015} and the references cited therein for the details and the proofs.

\subsection{Splines and collocation points}\label{subsec:splines}

We choose to apply a space of spline functions as ansatz space because of their advantageous properties, see e.g. \cite{Dahmen1998,PBP2002,Schumaker2007} for details on the theory of splines.

For a given interval $[a,b]$ we fix an equidistant knot sequence $T=\{t_i\}_{i=1}^{n+N}$ with $n$-fold knots at the interval end points $a=t_i$ for $1\le i \le n$ and $b=t_i$ for $N+1\le i \le N+n$. Moreover, we require that the knot sequence is strictly increasing inside the interval $(a,b)$, i.e. $t_{i} < t_{i+1}$ for $n \le i \le N$.

Then, an appropriate basis for our spline space is given by the B-spline functions $N_{i,n}$ of order $n$ which can be defined via the recursion formula
\begin{align*}
N_{i,1}(t) = \chi_{[t_{i}, t_{i+1}]}(t), \quad N_{i,n}(t) = (N_{i,n-1} * N_{i,1})(t),
\end{align*}
where $\chi_{[t_{i}, t_{i+1}]}$ is the characteristic function of the interval $[t_{i}, t_{i+1}] \subset \R$ and the convolution of two functions is defined as
$(f*g)(x):=\int_{\R} f(s) g(x-s) \, \dd s$.
Since we require that the ansatz functions are twice differentiable, we choose cubic splines, i.e. $n=4$.

In two dimensions the ansatz functions on a rectangular domain are obtained via a tensor ansatz and then are used as the $B_i$ in the previous subsection. The collocation points are chosen to coincide with the sequence of equidistant knots.
Since this leads to an underdetermined system of equations we use a \emph{not-a-knot condition} at the very but last knot at each interval end, i.e. we require that the spline function is three times continuously differentiable at this knot.
In other words, the restriction of the spline to the union of the two subintervals closest to each interval end is a cubic polynomial and the knot could be removed without changing the spline function.
This is a much simpler approach than the one used in the previous work~\cite[Section 4.2.3]{BHP2015} but provides approximately the same accuracy.


\section{Numerical solution of equations of Monge--Amp\`ere type for optical applications}\label{sec:MA_in_optics}

Next, we consider the particular difficulties that we have to overcome to efficiently solve the equations of Monge--Amp\`ere type that arise in the reflector and refractor problems.

\subsection{Boundary conditions}\label{subsec:boundary_condition}

The boundary conditions for both, the inverse reflector and refractor problems, are realized via a Picard-type iteration as similarly proposed by Froese~\cite[Section 3.4]{Froese2012}.

We assume that the light rays hitting the boundary of the optical surface also hit the boundary of the target, i.e. $\vec{z}(\partial\Omega)=\partial\Sigma$, see \cite[Section 4.5]{BHP2015} for the details.
This assumption is related to the edge ray principle, see e.g.~\cite[Appendix B]{WMB2005}.
Note that $\vec{z}$ also depends on the solution $\rho$ and its derivative $\GG \rho$.
In order to have a boundary condition which is easier to handle, we do not fix the target coordinate on the boundary but only its normal component. Since we do not know the normal component of the mapping $\vec{z}$ for the exact solution we proceed as follows:
For solving the nonlinear system of equations from our collocation method we use a Newton-type method producing iterations $\rho^k$, $k=1,2,...,n_{\max}$, starting with an initial guess $\rho^0$. We denote the corresponding mappings by $\vec{z}^{k}:=\vec{z}(\vec{x}',\rho^{k},\GG \rho^{k})$.
In the $k$th iteration we require that the outer normal of the mapping $\vec{z}^k$ of the current iteration and of the orthogonal projection of the mapping $\vec{z}^{k-1}$ of the last iteration onto the boundary coincide, i.e.
\begin{align*}
  \left(\vec{z}^k
      - \underset{\tilde{\vec{z}}\in\partial\Sigma}{\arg\min}\,
        \abslr{ \tilde{\vec{z}} - \vec{z}^{k-1} }^2
  \right) \cdot \nu(\vec{x}') = 0
    &&\text{for }\vec{x}'\in\partial\Omega,
\end{align*}
see \cite[Section 4.5]{BHP2015} (cf. also \cite[Section 3.3]{Froese2012}). The left-hand side is then used as function $G$ in Section~\ref{subsec:collocation}.

Since the last iteration is involved in the boundary condition the function $G$ changes in each iteration so that we solve different problems in successive steps.
In order to ensure the existence of a solution of the subproblems we follow the approach by Froese~\cite[Section 3.4]{Froese2012} and add a parameter $c$ in front of the right-hand side of the Monge--Amp\`ere equation~\eqref{eq:MA_refractor}, i.e. we replace $f$ by $cf$ where $c$ is an additional unknown in our equation. An additional constraint to compensate this new degree of freedom is discussed in Section~\ref{sec:size}.

\subsection{Ellipticity constraint}\label{sec:ellipticity}
For proofs of results for existence and uniqueness of a solution we require the equation of Monge--Amp\`ere type to be elliptic. In order to ensure ellipticity we manipulate the determinant in the same way as explained in~\cite[Section 4.4]{BHP2015} (cf. also \cite[Section 4.3]{Froese2012}): Let $\mathcal{W}=[\mathcal{W}_{i,j}]_{1\leq i,j\leq 2}\in\R^{2\times 2}$ be a matrix. For a penalty parameter $\lambda>0$ we define the modified determinant
\begin{equation*}
\begin{aligned}
  {\det}^+_{\lambda}\mathcal{W}
    :=& \max\{0,\mathcal{W}_{1,1}\}\max\{0,\mathcal{W}_{2,2}\} - \mathcal{W}_{1,2}^2\\
    &- \lambda\left[(\min\{0,\mathcal{W}_{1,1}\})^2 + (\min\{0,\mathcal{W}_{2,2}\})^2\right]
\end{aligned}
\end{equation*}
which we use instead of the determinant in the left-hand side of the Monge--Amp\`ere equation~\eqref{eq:MA_refractor}.
For an elliptic solution of this equation the left-hand side is exactly the same for the determinant and the modified determinant, see \cite[Lemma 4.2]{BHP2015}. Furthermore, each non-elliptic solution of \eqref{eq:MA_refractor} is not a solution of this equation, when the determinant is replaced by the modified determinant.

\subsection{Choice of the ``size'' of the refractor}\label{sec:size}
Up to now the refractor is at most uniquely determined up to its size. Therefore we define our initial guess $u_0$ of the problem appropriately and search for a solution $u$ of same size requiring that $\int_{\Omega} u  \, \dd s = \int_{\Omega} u_0  \, \dd s$ holds true, see also~\cite[Section 4.5]{BHP2015}.

Note that this condition is taken account of by the additional unknown $c$ introduced in Section~\ref{subsec:boundary_condition}.

\subsection{Total internal reflection}\label{sec:num_tir}
In case that $\kappa < 1$ it is possible that a ray of light exceeds the critical angle and total internal reflection occurs, such that this light ray does not reach the target.
Of course we know that this is not true for the solution, because we require that all light rays hit the target.
However during the iteration process of our nonlinear solver this phenomenon can appear.
If this is the case the argument of the square root in the definition of $\Phi$ in Section~\ref{subsec:snell} is negative at this position.

To overcome this instability we replace $\Phi(s)$ by its stabilized counterpart $\tilde{\Phi}(s) := s - \sqrt{\max\{0,\kappa^2+s^2-1\}}$.
Then the situation of total internal reflection is treated like the case when the light ray hits the surface exactly at the critical angle.
The refracted light ray most likely also misses the target and therefore this intermediate step cannot satisfy the Monge--Amp\`ere equation~\eqref{eq:MA_refractor} such that further iterations are performed.

If total internal reflection doesn't occur, which is the case we intend to have for our solution, we have $\Phi(s)=\tilde{\Phi}(s)$ and therefore obtain an equivalent problem.

\subsection{Nested iteration}\label{sect:NestedIteration}
The convergence of Newton-type methods sensitively depends on the choice of an initial guess that is close enough to the solution.
We apply a nested iteration strategy in order to largely increase the stability of the solver but also in order to accelerate the solution procedure. We start with a coarse grid for the spline surface and a blurred version of the image for the illumination pattern.

The blurring process is necessary because a coarse grid cannot produce a very detailed image on the target area.
For this reason we convolve the image, which is given as a raster graphic in our case, with a discrete version of the \emph{standard mollifier function} $\varphi(\vec{x}) := \exp(\nicefrac{-1}{(1-\abs{\vec{x}}^2)})$ if $\abs{\vec{x}}<1$ and zero otherwise, namely with
$
  \varphi_n(i,j):= \nicefrac{\varphi\left(2\frac{i}{n},2\frac{j}{n}\right)}
             {\sum_{r,s\in\Z}\varphi\left(2\frac{r}{n},2\frac{s}{n}\right)}
$
for $n\in\N$ and indices $i,j\in\Z$ for the pixel coordinates.

If our grid has $N\times N$ nodes we alternately increase the resolution $N$ of the grid and decrease the strength $n$ of blurring, i.e. we solve the problem for different pairs of $(N,n)$, see also \cite[Sections 4.3 and 5.2]{BHP2015}.

\subsection{Initial guess}\label{sec:initial_guess}
For the refractor problem we simply use the surface of a sphere with center at the position of the light source and a prescribed radius as initial guess.

For the reflector problem we start with a reflective surface producing a homogeneous illumination pattern on the target. We obtain this reflector by first using the method of supporting ellipsoids~\cite{KO1997,KO1998} and our collocation technique afterwards, see also~\cite[Section 5.2.3]{BHP2015}.

\subsection{Minimal gray value}\label{sec:min_gray}
The density function $g$ corresponds to the target illumination on $\Sigma$ and is given by $8$ bit digital grayscale images (integer gray values in the range from $0$ to $255$).
Since we divide by $g$ in right-hand side of the Monge--Amp\`ere equation~\eqref{eq:MA_refractor}, the function $g$ should be bounded away from zero.
To guarantee this lower bound we adjust the image and use the modified function
\begin{align}\label{eq:increas_gray_values}
  \tilde{g}(\vec{Z}):=g(\vec{Z})+\max\{0,L-\min_{\vec{Z'}\in\Sigma}g(\vec{Z'})\}
\end{align}
with $L\in\N$, see also~\cite[(5.9)]{BHP2015}. Numerical experiments show that the value $L=20$ leads to good results. In order to satisfy the energy conservation condition \eqref{eq:energy_conservation} the function $\tilde{g}$ needs to be scaled accordingly.


\section{Simulation results}\label{sec:results}

In this section we discuss some numerical simulation results
obtained by the collocation method for the inverse reflector and refractor problems.

\subsection{Lambertian radiator and target illumination}

\begin{figure}
  \centering
  \subfloat[Luminous intensity as a function on the angle $\theta$ between $z$-axis and direction of observation.]{
    \includegraphics[width=0.44\linewidth]{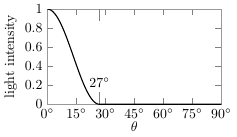}
  }
  \hspace{0.12\linewidth}
  \subfloat[Intensity as density function $f:\Omega\rightarrow\R^+$.]{
    \includegraphics[width=0.24\linewidth]{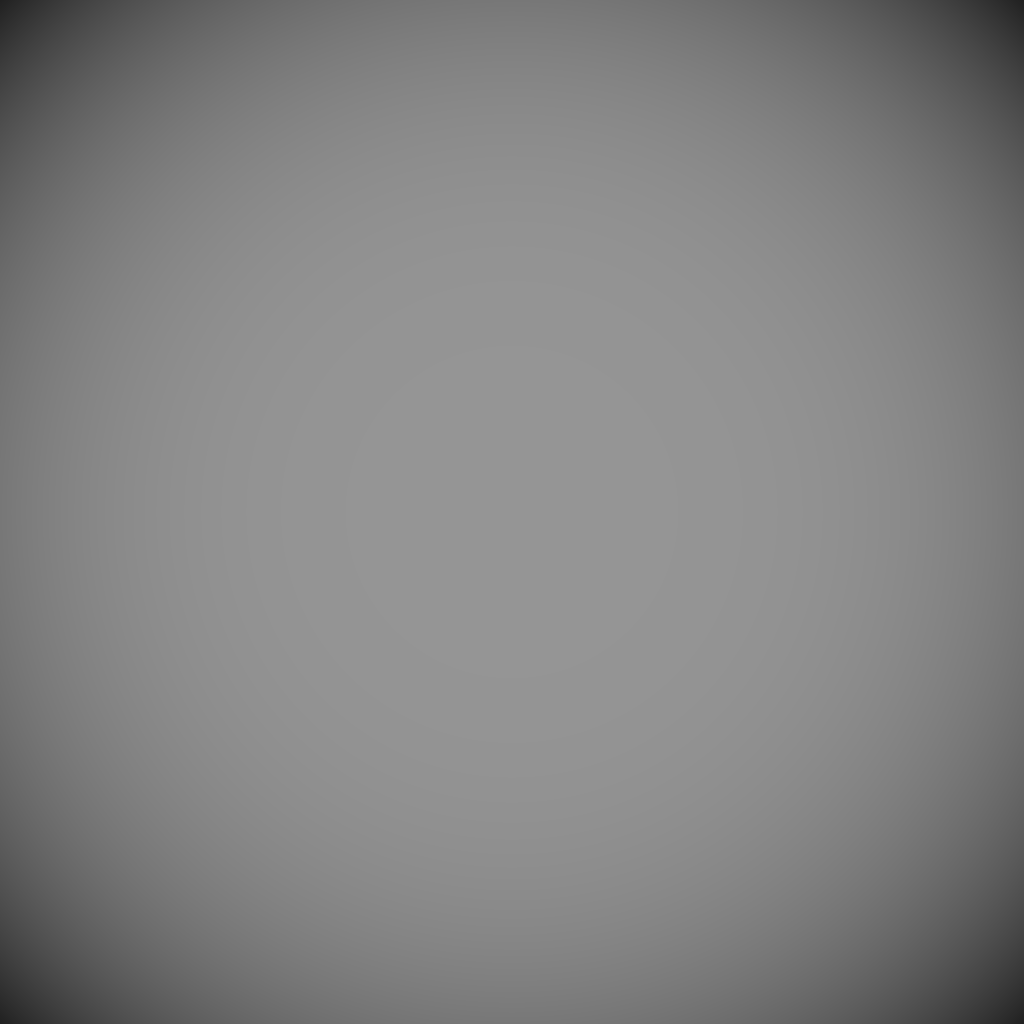}
  }
  \caption{Light emitting characteristics of the radiator of Lambertian type.}
  \label{fig:lightin}
\end{figure}

For both optical problems and all of our simulations we use
the domain $U=\{\vec{x}\in\cS^2\,:\, \vec{x}'\in (-\frac{3}{10},\frac{3}{10})^2\}$ and a light source with a Lambertian-type emission characteristics.
Its emitted luminous intensity $I(\theta)$ is rotationally symmetric, shows a fast decay and is proportional to $\cos{(\frac{20}{3}\theta)}$, where $\theta\in[0,\frac{3}{20}\pi]$ is the angle between the $z$-axis and the direction of observation.
Figure~\ref{fig:lightin} shows the emission density function $f$ depending on our two-dimensional parameter $\vec{x}'$ and on $\theta$.
We choose a light source with this characteristic because the maximum possible angular direction for our rectangular domain is about $\theta_{\max}=25^{\circ}$ and we therefore have a very low intensity at the edges of $\Omega$ to make the setting more difficult.

As desired target illumination patterns we chose four images with a variety of characteristics, i.e. many different patterns and features, see first row in Figure~\ref{fig:refractor_results1}.
The first three test images are taken from \cite{CVG}, while the fourth test image is our institute's logo.

\subsection{Geometrical setting and verification}

\begin{figure}
  \centering
  \includegraphics[width=0.6\linewidth]{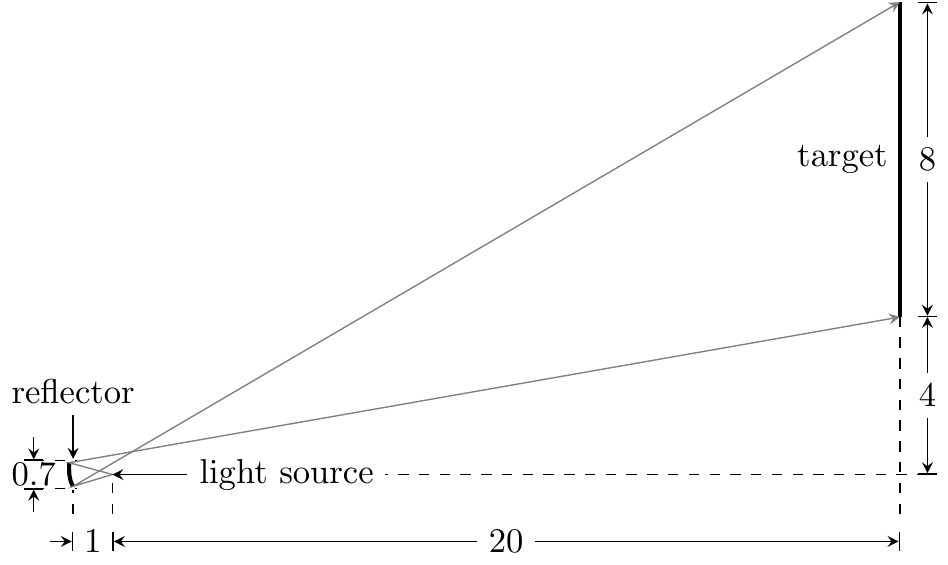}
  \caption{Geometrical setting of the examples for the reflector problem.}
  \label{fig:parametrization_mirror}
\end{figure}

Figure~\ref{fig:parametrization_mirror} shows our geometrical setting for the inverse reflector problem where the resulting reflectors are approximately of the size as in this figure. Here we have
$\Sigma=[4,12] \times [-4,4] \times \{20\}$.

\begin{figure}
  \centering
  \includegraphics[width=0.6\linewidth]{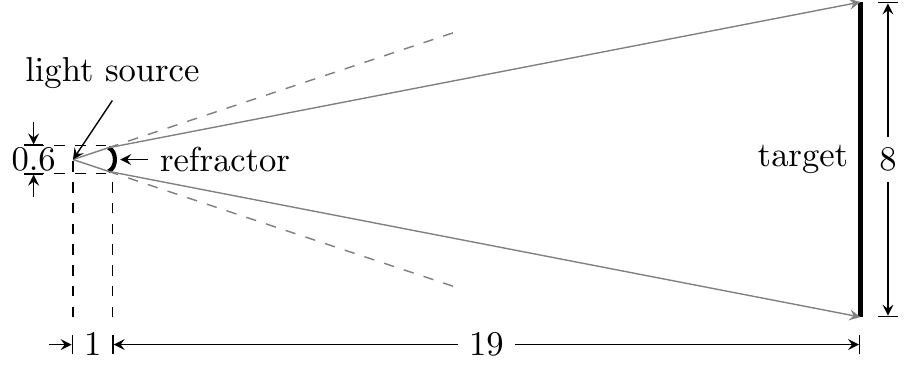}
  \caption{Geometrical setting of the examples for the refractor problem.}
  \label{fig:parametrization_lens}
\end{figure}

For the refractor problem the dimensions including the size of the optical surfaces are chosen very similarly to the case of the reflector problem to have a comparable situation, see Figure~\ref{fig:parametrization_lens}.
Here we use a part of the surface of a sphere with radius $0.5$ as initial guess, see also Section~\ref{sec:initial_guess}, and $\Sigma=[-4,4] \times [-4,4] \times \{ 20 \}$.
As refractive indices we use $n_1=\frac{3}{2}$ for the lens representing an average glass material and $n_2=1$ for the environment.

The calculated reflector or lens is verified using the ray tracing software POV-Ray~\cite{CFK+1991}.

\subsection{Choice of the parameters}

In the nested iteration we successively solve the nonlinear systems of equations for the following pairs $(N, n)$ of grid resolutions: $(16,163)$, $(31,163)$, $(31,55)$, $(61,55)$, $(61,19)$, $(121,19)$, $(121,7)$, $(241,7)$, $(241,3)$, and $(481,3)$, see Section~\ref{sect:NestedIteration} for the details.
The Newton-type method ends after at most $200$ iterations.

The regularization parameter in the modified determinant as defined in Section~\ref{sec:ellipticity} is set to $\lambda=10^3$ which turns out to be an appropriate choice for all examples.

\subsection{Results}

\begin{figure}[htb]
  \def\WIDTH{0.24\linewidth}
  \def\HSPACE{0.003\linewidth}
  \def\VSPACE{0.1pt}
  \centering

  \begin{tikzpicture}[overlay]
    \filldraw[black] (0.09,-4.66) rectangle (12.76, 4.84);
  \end{tikzpicture}
  \subfloat[``Boat'']{%
    \parbox{\WIDTH}{%
      \includegraphics[width=\linewidth]{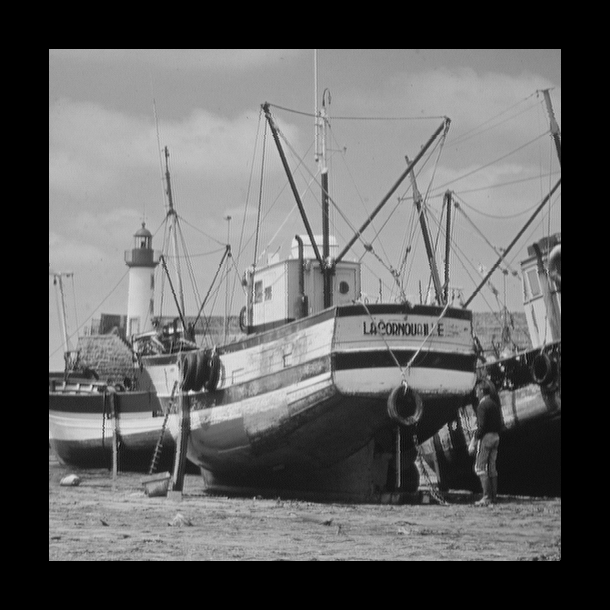}%
      \\\vspace{\VSPACE}%
      \includegraphics[width=\linewidth]{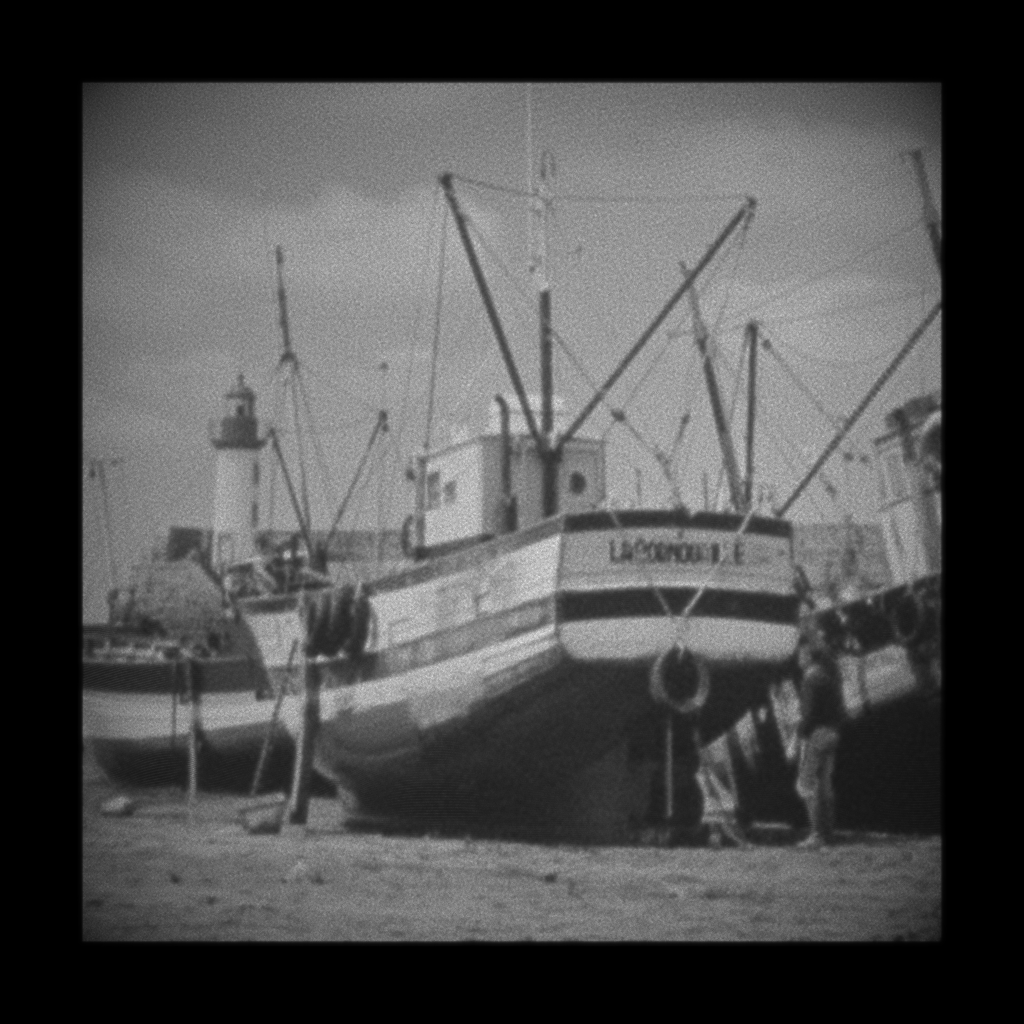}%
      \\\vspace{\VSPACE}%
      \includegraphics[width=\linewidth]{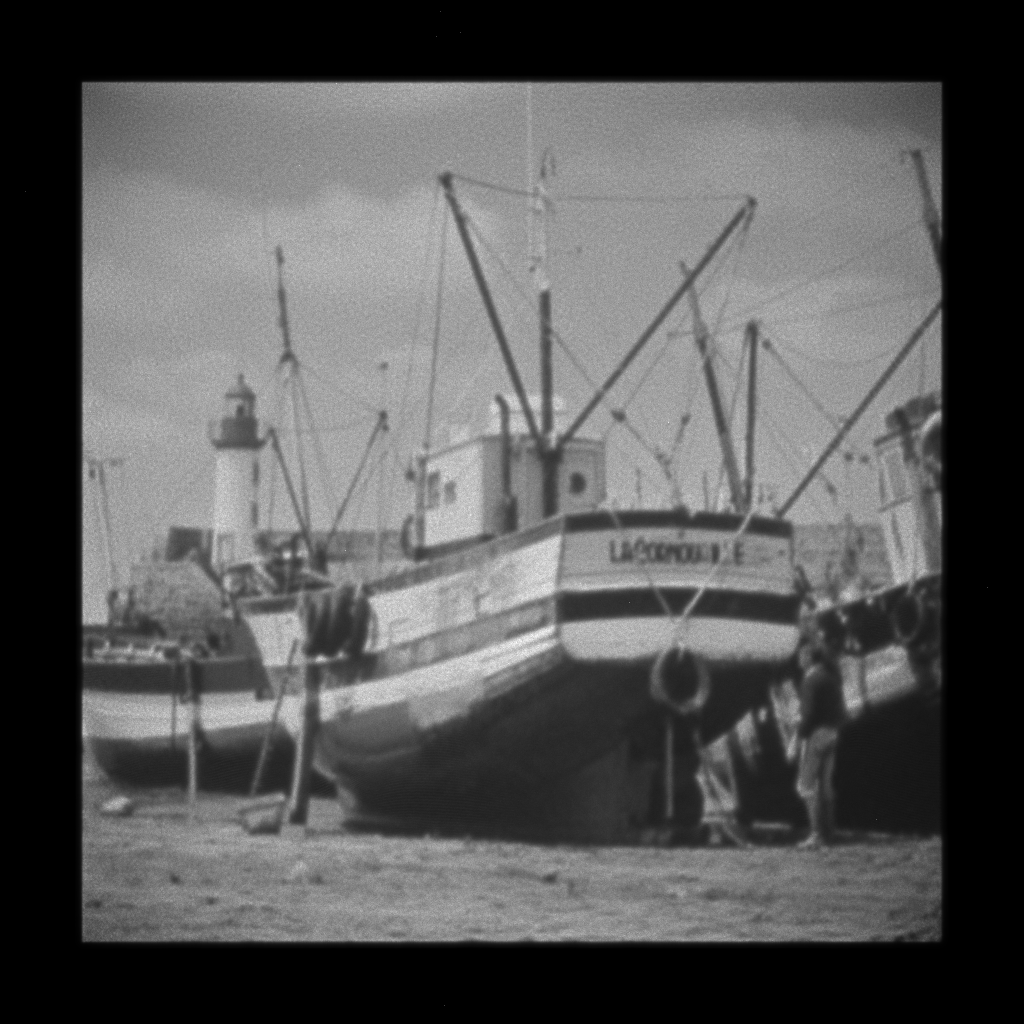}%
    }%
    \label{fig:reflector_results1_a}%
  }%
  \hspace{\HSPACE}%
  \subfloat[``Goldhill'']{%
    \parbox{\WIDTH}{%
      \includegraphics[width=\linewidth]{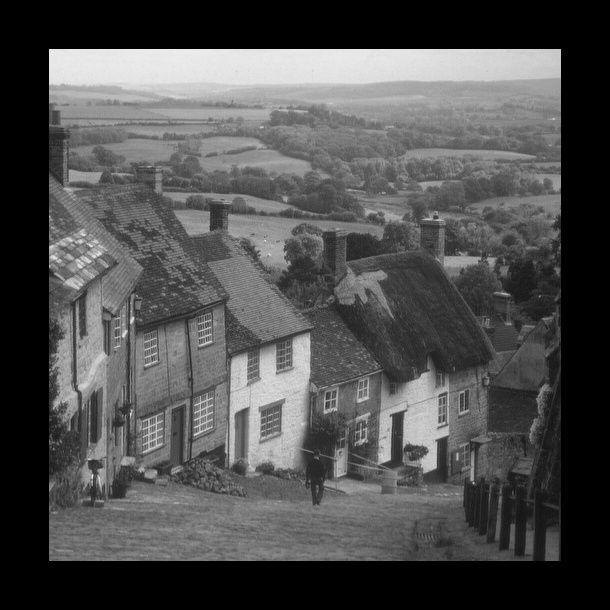}%
      \\\vspace{\VSPACE}%
      \includegraphics[width=\linewidth]{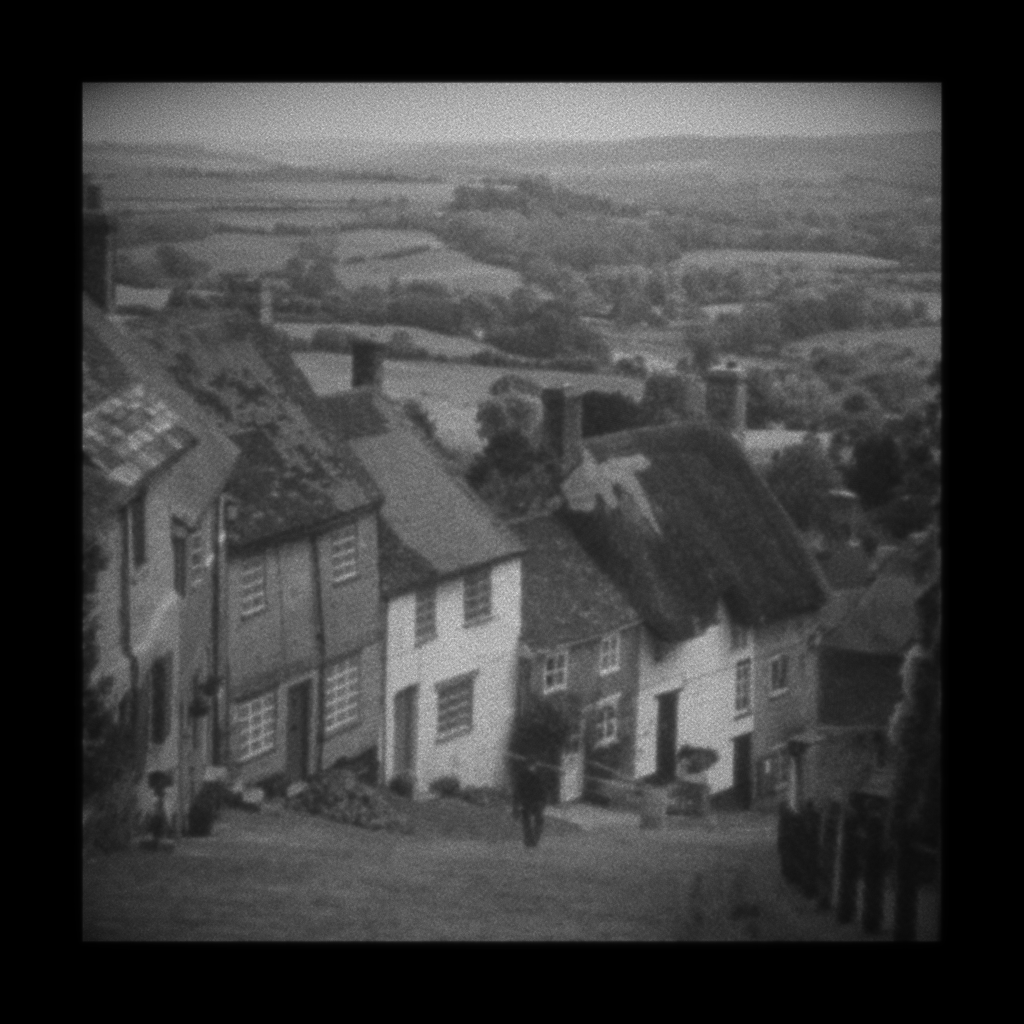}%
      \\\vspace{\VSPACE}%
      \includegraphics[width=\linewidth]{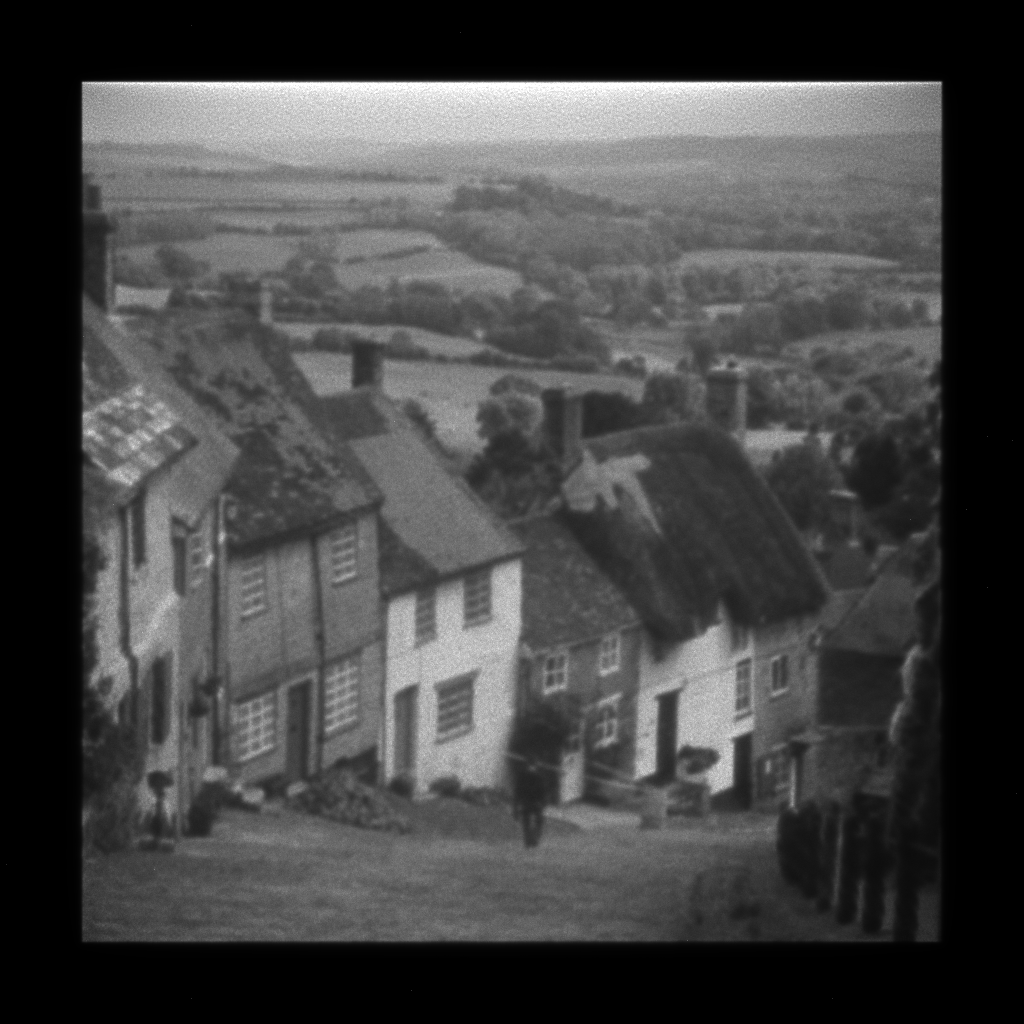}%
    }%
    \label{fig:reflector_results1_b}%
  }%
  \hspace{\HSPACE}%
  \subfloat[``Mandrill'']{%
    \parbox{\WIDTH}{%
      \includegraphics[width=\linewidth]{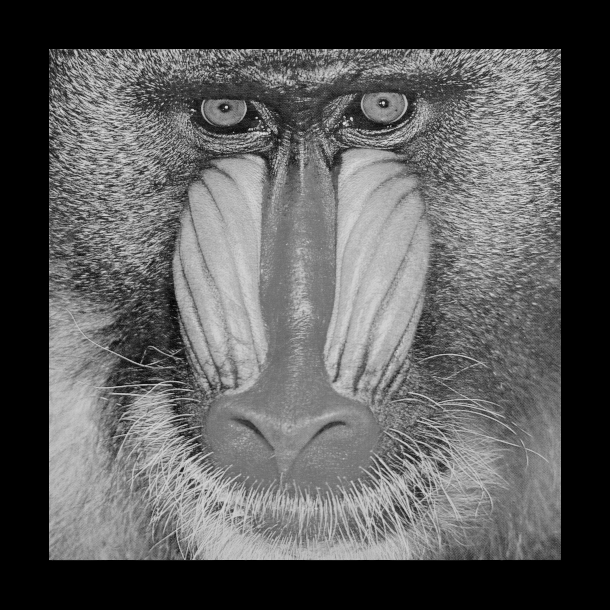}%
      \\\vspace{\VSPACE}%
      \includegraphics[width=\linewidth]{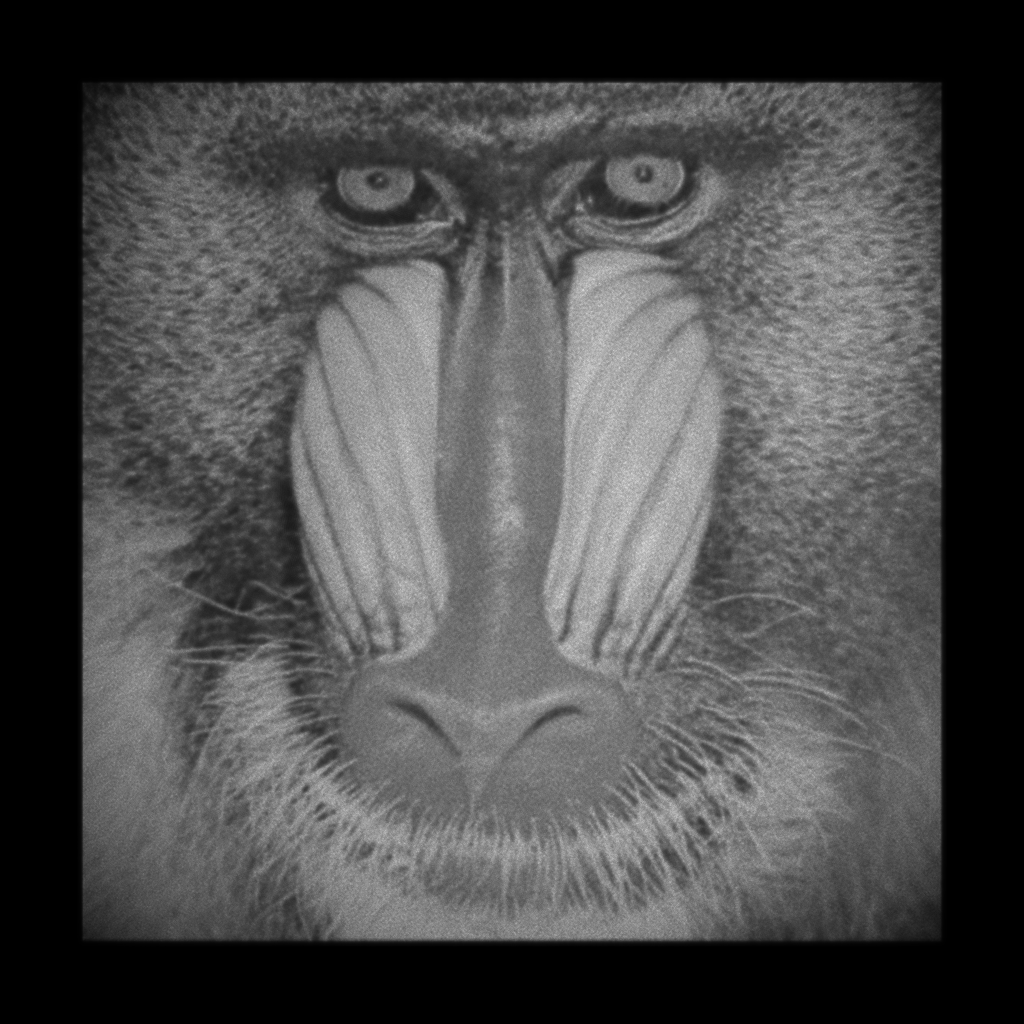}%
      \\\vspace{\VSPACE}%
      \includegraphics[width=\linewidth]{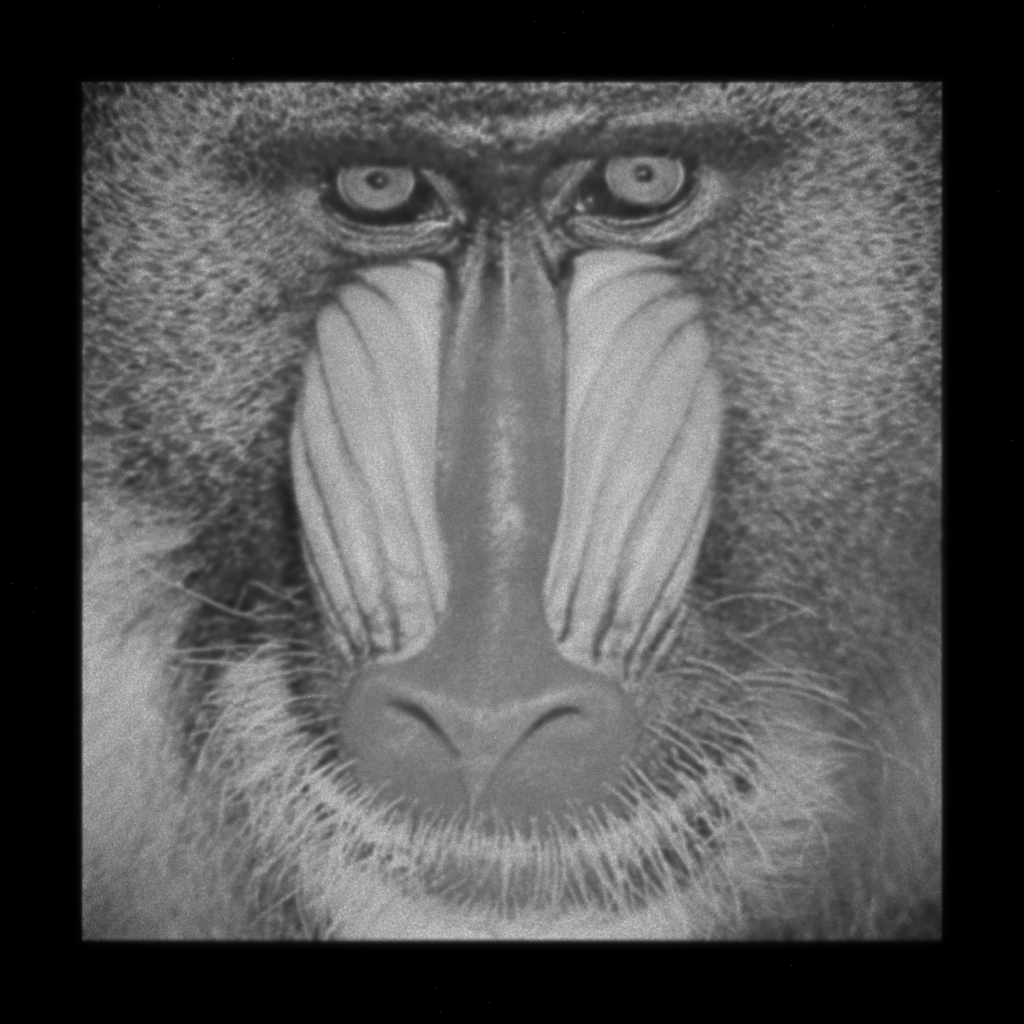}%
    }%
    \label{fig:reflector_results1_c}%
  }%
  \hspace{\HSPACE}%
  \subfloat[Institute's logo]{%
    \parbox{\WIDTH}{%
      \includegraphics[width=\linewidth]{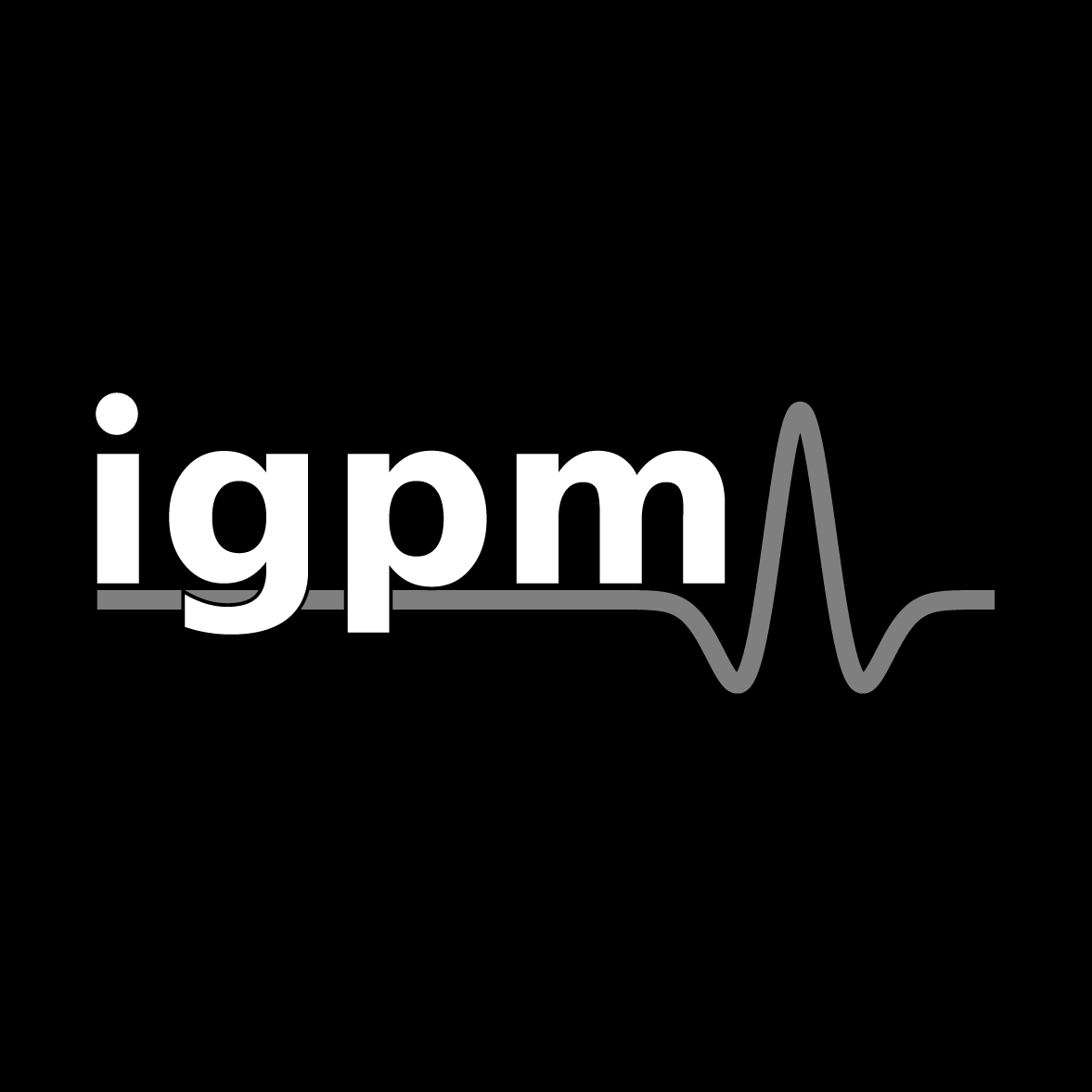}%
      \\\vspace{\VSPACE}%
      \includegraphics[width=\linewidth]{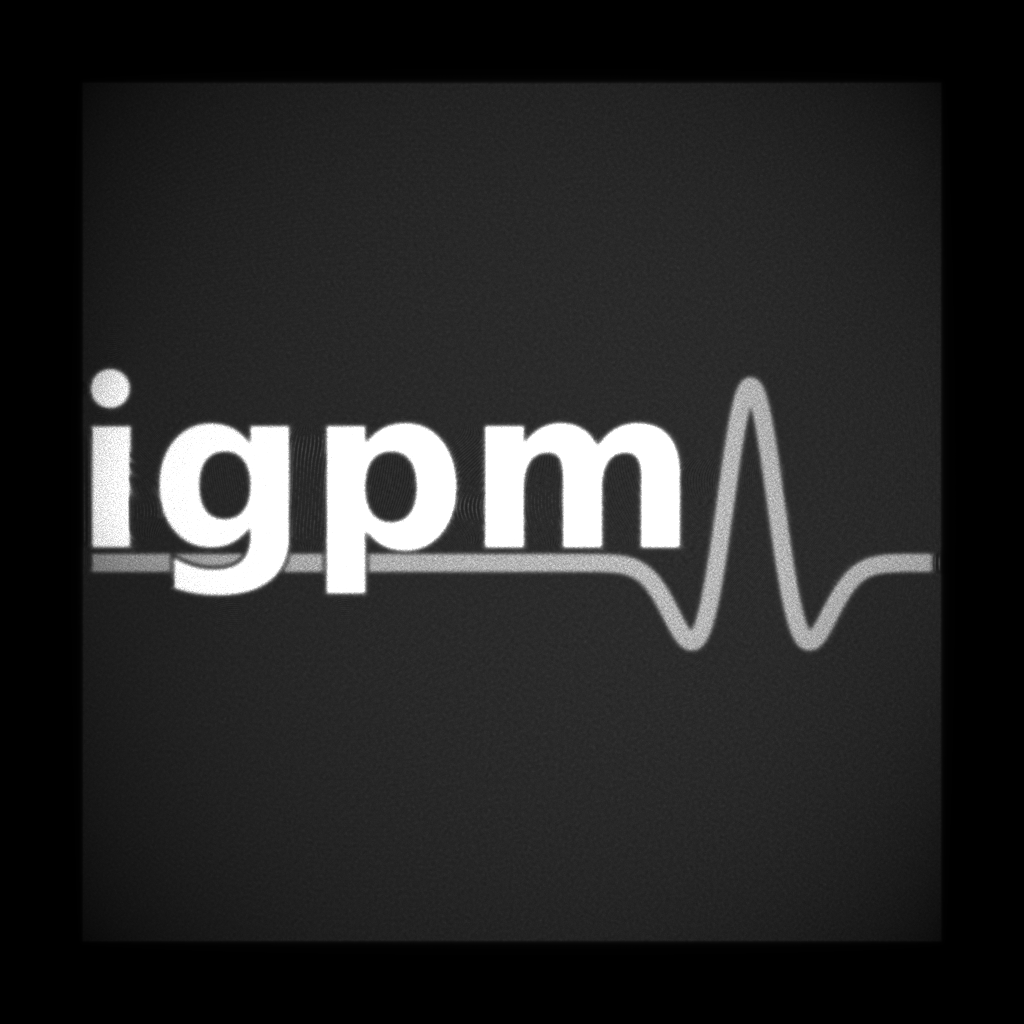}%
      \\\vspace{\VSPACE}%
      \includegraphics[width=\linewidth]{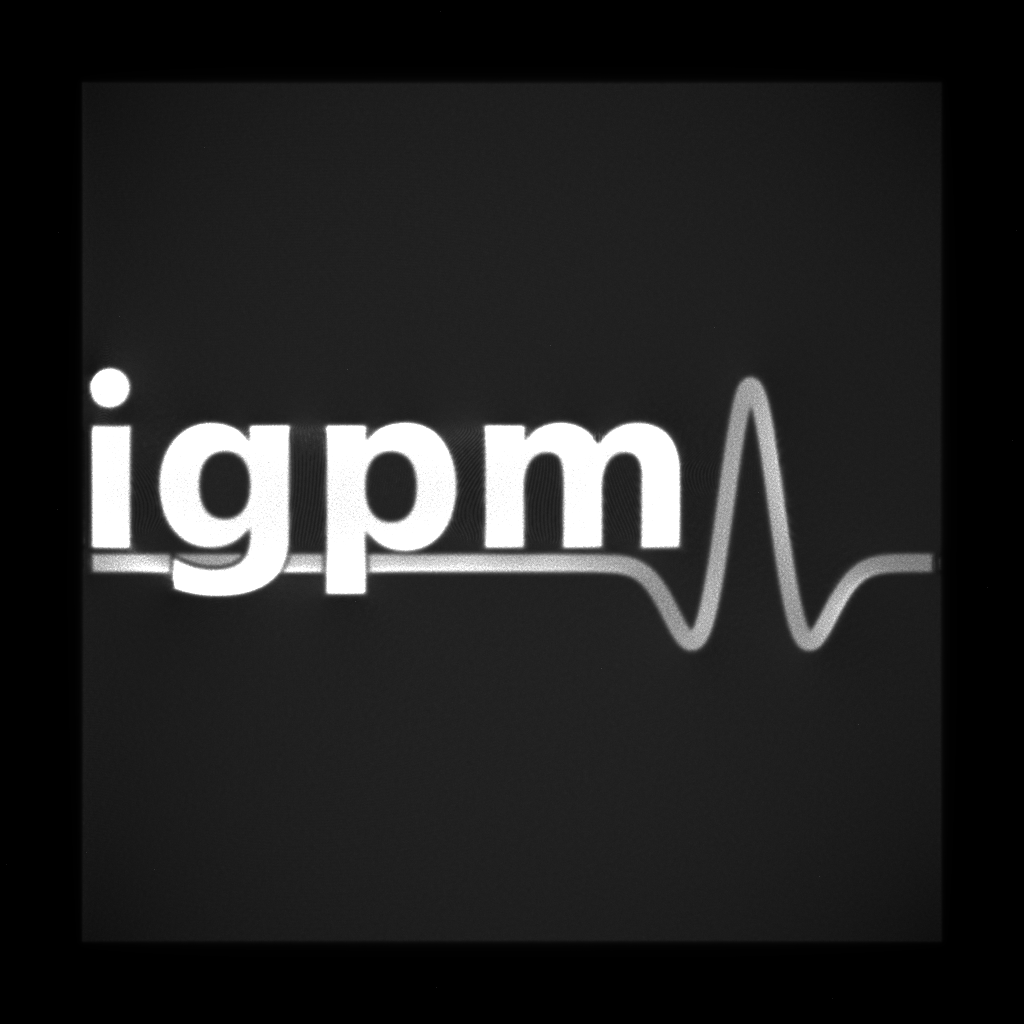}%
   }%
   \label{fig:reflector_results1_d}
  }
  \caption{Simulation results for three test images for the reflector and refractor problem.
           first row: desired distribution (original image, image sizes are $512\times 512$ pixel for the first three and $988\times 988$ pixel for the last image);
           second row: distribution after forward simulation by ray tracing for the reflector problem (result);
           third row: same as second row but for the refractor problem (result).}
  \label{fig:refractor_results1}
\end{figure}
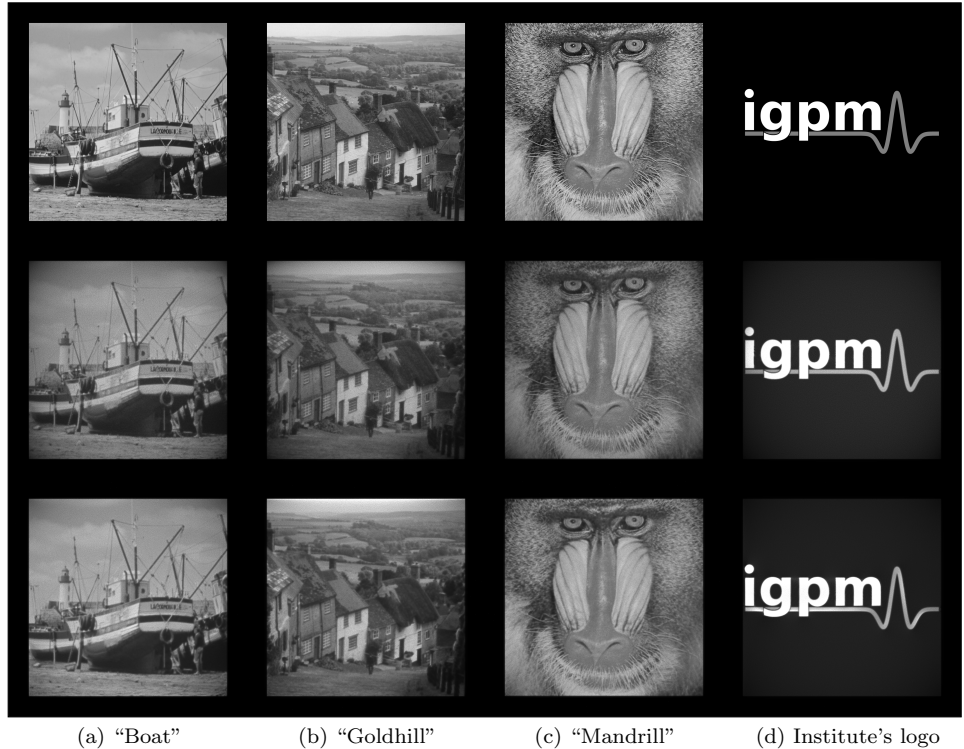

The results of the numerical simulations are depicted in Figure~\ref{fig:refractor_results1}. In the first row the
original test images are shown. The first three of them are chosen to examine different characteristics within the images, like thin straight lines and lettering as in the image ``Boat'', see Figure~\ref{fig:reflector_results1_a}.
Different patterns of high and low contrast are present in the image ``Goldhill'', see Figure~\ref{fig:reflector_results1_b}, in particular at the windows and roofs of the houses and the surrounding landscape in the background.
The image ``Mandrill'' in Figure~\ref{fig:reflector_results1_c}
shows the face of a monkey with a lot of fine details like the whiskers.
The fourth and most challenging of our test pictures is the logo of our institute in Figure~\ref{fig:reflector_results1_d} because it shows the highest possible contrast and contains jumps in the gray value from black to white.
The iteration counts and timings for the numerical experiments are given in Tables~\ref{tab:reflectors_a} and \ref{tab:reflectors_b}.

\begin{table}
  \centering
  \footnotesize
  \caption{Number of iterations of the Newton-type method for each of the ten nested iterations and overall computing time in seconds for the standard test images in Figure~\ref{fig:reflector_results1_a} ''Boat``, Figure~\ref{fig:reflector_results1_b} ''Goldhill``, and Figure~\ref{fig:reflector_results1_c} ''Mandrill``.}
    \begin{tabular}{rccc@{}p{3mm}@{}ccc}
      \hline
      Iterations & \multicolumn{3}{c}{refractor} && \multicolumn{3}{c}{reflector}\\
      \cline{2-4}  \cline{6-8}
      $(N,n)$
          & Boat
          & Goldhill
          & Mandrill
          &
          & Boat
          & Goldhill
          & Mandrill\\
      \hline
      $( 16,163)$           &  \phantom{0}70 & \phantom{0}65 &  \phantom{0}79 &&  \phantom{0}13 & \phantom{0}13 &  \phantom{0}11 \\
      $( 31,163)$           &  \phantom{0}13 & \phantom{0}13 &  \phantom{0}15 &&  \phantom{0}13 & \phantom{0}11 &  \phantom{0}11 \\
      $( 31,\phantom{0}55)$ &  \phantom{0}24 & \phantom{0}15 &  \phantom{0}15 &&  \phantom{0}13 & \phantom{0}13 &  \phantom{0}13 \\
      $( 61,\phantom{0}55)$ &  \phantom{0}43 & \phantom{0}15 &  \phantom{0}15 &&  \phantom{0}13 & \phantom{0}13 &  \phantom{0}13 \\
      $( 61,\phantom{0}19)$ &  \phantom{0}35 & \phantom{0}44 &  \phantom{0}37 &&  \phantom{0}13 & \phantom{0}13 &  \phantom{0}13 \\
      $(121,\phantom{0}19)$ &  \phantom{0}41 & \phantom{0}32 &  \phantom{0}43 &&  \phantom{0}13 & \phantom{0}13 &  \phantom{0}13 \\
      $(121,\phantom{00}7)$ &  \phantom{0}47 & \phantom{0}41 &  \phantom{0}42 &&  \phantom{0}13 & \phantom{0}18 &  \phantom{0}18 \\
      $(241,\phantom{00}7)$ &  \phantom{0}46 & \phantom{0}38 &  \phantom{0}40 &&  \phantom{0}13 & \phantom{0}15 &  \phantom{0}13 \\
      $(241,\phantom{00}3)$ &  \phantom{0}39 & \phantom{0}41 &  \phantom{0}54 &&  \phantom{0}15 & \phantom{0}15 &  \phantom{0}26 \\
      $(481,\phantom{00}3)$ &  \phantom{0}38 & \phantom{0}36 &  \phantom{0}45 &&  \phantom{0}15 & \phantom{0}15 &  \phantom{0}24 \\
      \hline
      \multicolumn{1}{l}{Time / s} & 227 & 210 & 234 && \phantom{0}90 & \phantom{0}90& 133\\
      \hline
    \end{tabular}
    \label{tab:reflectors_a}
\end{table}

\begin{table}
  \centering
  \footnotesize
  \caption{Number of iterations of the Newton-type method for each of the ten nested iterations and overall computing time in seconds for the institute's logo in Figure~\ref{fig:reflector_results1_d}.}
    \begin{tabular}{rc@{}p{3mm}@{}c}
      \hline
      Iterations & refractor && reflector\\
      \cline{2-2}\cline{4-4}
      $(N,n)$
          & Institute's logo
          && Institute's logo\\
      \hline
      $( 21,          100)$ &  \phantom{00}33 && \phantom{0}54\\
      $( 41,          100)$ &  \phantom{00}13 && \phantom{0}11\\
      $( 41,          100)$ &  \phantom{00}11 && \phantom{0}11\\
      $( 81,          100)$ &  \phantom{00}13 && \phantom{0}11\\
      $( 81,\phantom{0}73)$ &  \phantom{00}54 && \phantom{0}19\\
      $(161,\phantom{0}73)$ &  \phantom{00}35 && \phantom{0}13\\
      $(161,\phantom{0}25)$ &  \phantom{0}200 && \phantom{0}90\\
      $(321,\phantom{0}25)$ &  \phantom{00}66 && \phantom{0}20\\
      $(321,\phantom{00}9)$ &  \phantom{0}200 &&           155\\
      $(641,\phantom{00}9)$ &  \phantom{00}59 && \phantom{0}22\\
      \hline
      \multicolumn{1}{l}{Time / s} & 1390 && 928\\
      \hline
    \end{tabular}
    \label{tab:reflectors_b}
\end{table}

First, we notice that for a given original image the output images obtained by forward simulation for the reflector and refractor problem look very similar.
In comparison to the original images the output images are slightly blurred and have a little less contrast but visually they only differ locally at very few locations.
Major deviations can be observed in the background of the institute's logo which is not completely black after the forward simulation of the mirror and the lens.
This is because of the minimal gray value needed to avoid the division by zero, see Section~\ref{sec:min_gray}.

We see that all of these characteristics of the first three test images are well preserved by our method.
The computing time for the refractor is approximately twice as long as for the reflector but still acceptable with about $4$ minutes.

For the fourth test image we had to adjust the parameters in the nested iteration process to handle the sharp edges and work with a finer grid, see Table~\ref{tab:reflectors_b}.
We also raised the minimal gray value in Section~\ref{sec:min_gray} from $20$ to $30$ obtaining a proportion between black and white of $1:9.5$.
These parameters lead to results showing also a very sharp logo for both the inverse refractor and reflector problems.
Note that nevertheless in two stages of the nested iteration for the refractor problem the quasi-Newton method was stopped because the maximal number of iterations was reached without meeting the required tolerances, see Table~\ref{tab:reflectors_b}. This happened only for two intermediate steps of the nested iteration process while we observe convergence in the last iteration, which shows us that this does not affect the overall method.
In the case of the refractor problem the gray line below the letters is irregularly illuminated and slightly too bright.
Nevertheless, the shape of this line is reproduced very precisely.

\begin{figure}
  \centering
  \subfloat[Refractor surface in correct geometrical position (overview)]{
    \includegraphics[width=0.4\linewidth]{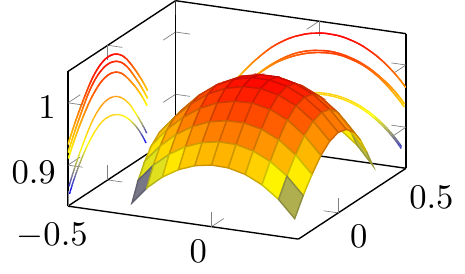}
  }
  \hspace{0.05\linewidth}
  \subfloat[High-frequency components of the refractor (fine structure).]{
    \includegraphics[width=0.4\linewidth]{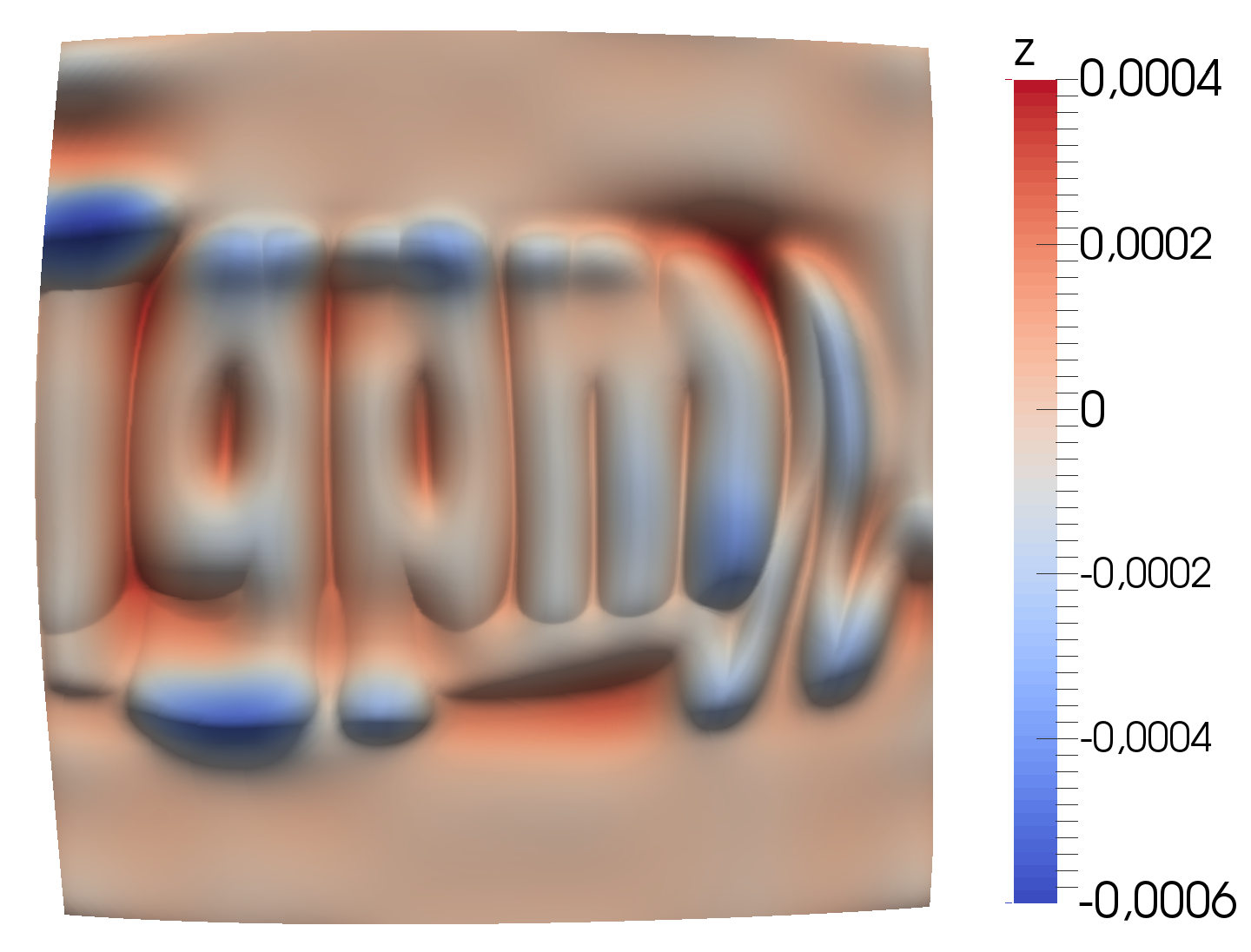}
  }
  \caption{Outer refractor surface for projecting our institute's logo.}
  \label{fig:refractor_results2}
\end{figure}

The optically active surface of the lens for the projection of the institute's logo is displayed in Figure~\ref{fig:refractor_results2}.
Note that the characters used in the logo can be recognized on the surface.
We observe that they cover about the half of the lens' surface while this is not the case in the original image.
Of course this is what we expect because we want to redirect a maximal amount of incoming light onto these letters.

\section{Summary and outlook}\label{sec:outlook}

For the efficient and stable solution of the inverse reflector and refractor problems we propose a numerical B-spline collocation method which is applied to the formulation of the inverse optical problems as partial differential equations of Monge--Amp\`ere type and appropriate boundary conditions.
Several challenges for the construction of a stable numerical solution method have been met, e.g. we detailed how to enforce ellipticity constraints to ensure uniqueness of the solution and how to handle the involved boundary conditions.
A nested iteration approach simultaneously considerably improves the convergence behavior and speeds up the numerical procedure.

For the inverse refractor problem our algorithm provides a reliable and fast method to compute one of the two surfaces of the lens under the assumption of a point-shaped light source.
Shaping the second surface of the lens, e.g. to minimize Fresnel losses, and exploring possible solution strategies for the problem for extended real light sources are topics of upcoming research.


\section*{Acknowledgments}

The authors are deeply indebted to Professor Dr. Wolfgang Dahmen for many fruitful and inspiring discussions on the topic of solving equations of Monge--Amp\`ere type.
We thank Elisa Friebel, Silke Glas, and Gudula K\"ammer for proofreading the manuscript.



\end{document}